\documentclass[aps,prl,twocolumn,groupedaddress]{revtex4}
\usepackage{graphicx}
\usepackage{amsmath,amsfonts,amssymb}
\usepackage{color}
\begin{document}

\title{Spontaneously Induced General Relativity: \\
Holographic Interior for Reissner-Nordstrom Exterior}

\author{Aharon Davidson}
\email{davidson@bgu.ac.il}
\homepage[Homepage: ]{http://www.bgu.ac.il/~davidson}
\author{Ben Yellin}
\email{yellinb@bgu.ac.il}

\affiliation{Physics Department, Ben-Gurion University of the Negev,
Beer-Sheva 84105, Israel}

\date{\today}

\begin{abstract}
If general relativity is spontaneously induced, that is if the reciprocal
Newton constant serves as a VEV, the electrically charged black hole
limit is governed by a Davidson-Gurwich phase transition which occurs
precisely at the would have been outer horizon.
The transition profile which connects the exterior Reissner-Nordstrom
solution with the novel interior is analytically derived.
The inner core is characterized by a vanishing spatial volume and
constant surface gravity, and in some respects, resembles a maximally
stretched horizon.
The Komar mass residing inside any concentric interior sphere is
proportional to the surface area of that sphere, and consequently,
is non-negative definite and furthermore non-singular at the origin.
The Kruskal structure is recovered, admitting the exact Hawking imaginary
time periodicity, but unconventionally, with the conic defect defused at the
origin.
The corresponding holographic entropy packing locally saturates the
't Hooft-Susskind-Bousso holographic bound, thus making the core
Nature's ultimate information storage.

\end{abstract}


\maketitle

\section{Introduction}
Bekenstein-Hawking area entropy \cite{HB}, which plays a central
role in black hole thermodynamics, has given rise to the speculative
idea that no physical degrees of freedom reside within the interior
of a black hole.
Such an idea is theoretically backed by the fact that neither the
Gibbons-Hawking \cite{GH} Euclidean path integral derivation, nor
the more locally oriented Wald's \cite{Wald} derivation, make actually
use of the black hole interior.
One may thus conclude that, as far as entropy packing is concerned,
the interior of a black hole is apparently superfluous, so that the black
hole degrees of freedom, whatever they are, live on or just above the
outer horizon. 
The deal being one bit of information per a quarter of Planck area of
the horizon surface \cite{bit}.
The apparent inconsistency between the horizon as a physical entity,
the residence \cite{stretch} of the black hole degrees of freedom, and
as the mere point of no return for all in-falling matter, has ignited a
well advertised debate in the physical society.
The black hole area entropy formula has inspired the so-called
holographic principle.
The latter asserts that all of information contained in some region of
space can be represented as a 'hologram' on the boundary of that region.
It furthermore puts a universal purely geometrical bound, saturated
by Bekenstein-Hawking area entropy, on the amount of entropy
stored within that region, namely
\begin{equation}
	S\leq \frac{A}{4G}~,
	\label{bound}
\end{equation}
where $A$ denotes the area of the closed spacial boundary, $G$
is Newton's constant, and $\hbar=c=k_B=1$.
The holographic principle, primarily introduced by 't Hooft \cite{tHooft},
attempting to resolve the black hole information paradox, has been
further developed by Susskind \cite{Susskind} to deal with black hole
complementarity, and has eventually acquired a covariant generalization
by Bousso \cite{Bousso}.
The holographic principle is recently gaining a major theoretical support
from the AdS/CFT duality \cite{AdsCFT}.

It is commonly believed that general relativity is not necessarily the
ultimate theory of gravity.
If it is not the fundamental, but rather (say) a spontaneously induced
theory of gravity, with $G^{-1}$ treated as a VEV, the black hole limit
has been shown \cite{Essay} to be governed by a phase transition which
occurs precisely at the would have been event horizon.
Recall that the idea of horizon phase transition \cite{PhaseTransition}
is not new (and in a similar category are black stars \cite{BlackStar}
and stringy fuzzballs \cite{fuzzballs}).
Whereas the general relativistic exterior black hole solution is fully
recovered, it serendipitously connects now, by means of a smooth self
similar transition profile, with a novel holographic core.
This core is characterized, among other things, by a vanishing spatial
volume, a crucial feature for black hole physics.
It is in this context of spontaneously induced general relativity that
the first local realization of maximal entropy packing has been
demonstrated \cite{DG}.
To be a bit more specific, sticking momentarily to spherical symmetry,
it has been shown that associated with any inner sphere of circumferential
radius $r$ is the total purely geometrical universal entropy
\begin{equation}
	S(r)=\frac{\pi r^2}{G}~,
	\label{Sr}
\end{equation}
which saturates the holographic bound layer by layer.
The accompanying Komar mass \cite{Komar}, as well as Weinberg's material
energy \cite{Weinberg}, are notably non-singular, namely
\begin{equation}
	M(r)=M(h)\frac{r^2}{h^2}~,
	\label{Mr}
\end{equation}
where $M(h)$ denotes the overall mass calculated at $r=h$, and $A=\pi h^2$
is the horizon area.
Such a Komar mass distribution appears to be intimately related to, and thus
as fundamental as, the entropy distribution itself.
In addition, the fact that the corresponding invariant spatial volume
$V(r)\rightarrow 0$ for every $r\leq h$ can explain why the black
hole entropy, unlike in any other macroscopic system,  is not proportional
to the volume of the system.
Rather than envision bits of information evenly spread on the horizon surface,
they may actually inhabit, universally and holographically in an onion-like
manner, the entire black hole interior.

In this paper, by providing a holographic interior for the Reissner-Nordstrom
exterior, we extend the Davidson-Gurwich analysis\cite{DG}.
Our paper is organized as follows.
We begin by motivating and then introducing the action which governs the
spontaneously induces general relativity, and derive the associated gravitational
and scalar field equations (section II).
These equations do not seem to admit a generic analytical solution, so we start
by deriving their asymptotic behavior for the static spherically symmetric case,
focusing on the deviation from the Reissner-Nordstrom background (section III).
The asymptotic expansion is then used as a boundary condition for numerically
plotting the various functions floating around, and to get a first glimpse into
the characteristic phase transition which is developed at the would have
been black hole horizon (section IV).
At this stage, one can already appreciate the vanishing invariant volume of the
novel interior core.
Consequently, by systematically getting rid of the negligible terms in the field
equations, we derive the approximate analytic solution of the core metric
(section V).
The in-out self-similar transition profile is analytically calculated (section
VI), allowing us to finally fix the left over parameters of the inner solution by
means of the mass $M$ and the elecric charge $Q$.
Various aspects of the inner metric, such as vanishing volume, light cones, Kruskal
structure, and singularity issues are discussed in section VII.
Finally, owing to the recovery of the exact Hawking imaginary time periodicity
(by defusing a conic singularity near the origin) and the emerging of a 
characteristic non-singular Komar mass function (section VIII), we re-formulate
the holographic entropy packing, emphasizing its universal geometric structure
(section IX).

\section{Action and Field equations}
The simplest theory which accounts for the coupling of electromagnetism
to spontaneously induced general relativity is given by the action
\begin{equation}
	{\cal I}=-\frac{1}{16\pi}\int\left(\phi R+V(\phi)
	+F^{\mu\nu} F_{\mu\nu}\right)\sqrt{-g}~d^4x ~.
	\label{eq:basicAction}
\end{equation}
The role of the scalar potential is to allow the conformally coupled
Brans-Dicke \cite{BD} scalar field $\phi(x)$ to develop the vacuum
expectation value (VEV)
\begin{equation}
	\langle \phi \rangle=\frac{1}{G}~.
\end{equation}
Several remarks are in order:

\noindent (i) The scalar field is kept electrically neutral, and furthermore
does not couple directly to electromagnetism.
This defines the Jordan frame, rather than the Einstein frame, to be the
physical one.
Our main conclusions turn out, however, to be frame independent.

\noindent (ii) On simplicity grounds, a kinetic scalar term has not been
introduced.
This, as we shall see, does not make the scalar field non-dynamical.
Adding a kinetic term is always a viable option though, with minor effects
on the inner metric.

\noindent (iii) The specific choice of the double well scalar potential
\begin{equation}
	V(\phi)=\frac{3}{2a}\left(\phi-\frac{1}{G}\right)^2~,
	\label{V}
\end{equation}
makes the theory fully equivalent to a simple $f(R)$ gravity \cite{f(R)}
theory, namely
\begin{equation}
	{\cal I}=-\frac{1}{16\pi}\int\left(\frac{1}{G}R-\frac{a}{6}R^2
	+F^{\mu\nu} F_{\mu\nu}\right)\sqrt{-g}~d^4x~.
\end{equation}
In particular, stability $\acute{\text{a}}$ la Sotiriou-Faraoni \cite{stability}
is guaranteed by construction for $a>0$.
The value of $a$ can be made as small as necessary to be compatible
with Solar System tests.

Varying the action eq.(\ref{eq:basicAction}) with respect to the three dynamical
fields $g_{\mu\nu},A_\mu,\phi$ leads respectively to the following equations
of motion
\begin{eqnarray}
	&& \phi G_{\mu \nu} +\phi_{;\mu \nu} - g_{\mu \nu} \square \phi \notag\\
	&&\quad=\frac{V(\phi)}{2}g_{\mu \nu}
	-2F_{\mu\alpha} F_{\nu \beta}g^{\alpha\beta}
	+ \frac{F^2}{2} g_{\mu \nu} ~,
	\label{eq:gFieldEq} \\
	&& F^{\mu \nu}_{~;\nu} =0 ~,
	\label{eq:AFieldEq}\\
	&& R+\frac{dV(\phi)}{d\phi} =0 ~.
	\label{eq:phiFieldEq}
\end{eqnarray}
By tracing the gravitational field eqs.(\ref{eq:gFieldEq}), and then
substituting the resulting Ricci scalar into eq.(\ref{eq:phiFieldEq}), one
can extract the associated Klein Gordon equation
\begin{equation}
	\square \phi =\frac{1}{3}\left(
	\phi\frac{dV(\phi)}{d\phi}-2V(\phi)\right)
	\equiv  \frac{dV_{eff}(\phi)}{d\phi}~.
	\label{eq:KG}
\end{equation}
The evolution of the scalar field is thus governed by the effective potential
\begin{equation}
	V_{eff}(\phi)=\frac{1}{2aG}\left(\phi-\frac{1}{G}\right)^2+ const~.
	\label{Veff}
\end{equation}
The similarity between the two potentials $V(\phi)$ and $V_{eff}(\phi)$ is
not generic.

At this stage, our interest lies with the static spherically symmetric case,
with the corresponding line element taking the conventional form
\begin{equation}\label{eq:basicMetric}
 	ds^2=-e^{\nu(r)}dt^2+e^{\lambda(r)}dr^2 + r^2 d \Omega^2 ~.
\end{equation}
The only non-vanishing entries of the electromagnetic tensor are
\begin{equation}\label{eq:emTensor}
	 F_{tr}=-F_{rt}=E(r) ~.
\end{equation}
First, we solve the associated Maxwell equation
\begin{equation}
	E'(r)+\left(\frac{2}{r}-\frac{ \nu '(r)+ \lambda '(r)}{2}\right)E(r)=0~,
\end{equation}
whose straight forward solution is given by
\begin{equation}\
	 E(r)=\frac{Q}{r^2} e^{\frac{\lambda (r)+\nu (r)}{2}}~,
	 \label{eq:electic-field}
\end{equation}
with $Q$ denoting the electric charge.
Next, we substitute eq.(\ref{eq:electic-field}) into the three independent
gravitational and scalar field equations, which can then be reorganized
in the master form
\begin{equation}
	\phi ^{\prime\prime }-\frac{1}{2}\left(\nu'
	+\lambda'\right)\left(\phi '+\frac{2}{r}\phi \right)=0~,
	\label{eq:FieldEqI}
\end{equation}
\begin{eqnarray}
	 &&  \phi''+\frac{1}{2}\left(\nu'-\lambda'\right)
	 \left(\phi'-\frac{2}{r}\phi\right)
	 -\frac{2 }{r^2}\left(1-e^{\lambda }\right) \phi \notag \\
 	&&\quad\quad =\frac{3e^{\lambda}}{2a}\left(\phi 
	-\frac{1}{G}\right)\left(\phi +\frac{1}{3G}\right)
	+2e^{\lambda}\frac{Q^2}{ r^4}  ~,
	\label{eq:FieldEqIII}
\end{eqnarray}
\begin{equation} 
	\phi''  +\left(\frac{2 }{r}+\frac{\nu'
	-\lambda'}{2}\right) \phi '
	=\frac{e^{\lambda}}{aG} 
	\left(\phi-\frac{1 }{G}\right) ~.
 	\label{eq:FieldEqII}
\end{equation}
One can easily verify that associated with the vacuum solution
$\displaystyle{\phi(r)=\frac{1}{G}}$ is the Reissner Nordstrom (RN) black
hole metric
\begin{equation}
	e^{\nu(r)}=e^{-\lambda(r)}=1-\frac{2G M}{r}+\frac{G Q^2}{r^2}~.
\end{equation}
However, unlike in general relativity, the Reissner-Nordstrom solution,
which we hereby tag with some $\epsilon=0$, is now accompanied by
a general class of asymptotically flat $\epsilon\neq 0$ solutions.
In this language, our paper is mainly devoted to the unfamiliar physics
encountered at the $\epsilon\rightarrow 0$ limit.

\section{Asymptotic behavior}

A general analytic solution of our field equations is still at large.
Alternatively, we adopt the strategy to use an asymptotically flat perturbation
around the RN solution as a boundary condition for the numerical solution of
the field equations.
Needless to say, the numeric solution by itself is not our final goal,
but the resulting graphs will give us the first clue regarding the structure
of the phase transition awaiting ahead.
Consider thus a perturbative solution of the general form
\begin{align}
	& \phi (r)= \frac{1}{G}\left(1+s \phi _1(r)\right) ~,\\
	& \lambda (r) = -\log\left(1-\frac{2 G M}{r}
	+\frac{G  Q^2}{r^2}+s L_1(r)\right)~,
	\label{eq:lambda-basic-form}\\
	& \nu (r) = \log\left(1-\frac{2GM}{r}
	+\frac{G  Q^2}{r^2}+s N_1(r)\right)~,
	\label{eq:nu-basic-form}
\end{align}
where the constant $s$, which can be interpreted as the scalar charge,
serves as our small expansion parameter.
Naively, one may expect general relativity to be fully recovered at the
limit $s\rightarrow 0$, but as we are about to see, this is not necessarily
the case.
Of particular interest for us is the decoupled linear differential equation 
\begin{eqnarray}
	 &&\left(1-\frac{2GM }{r}+\frac{G Q^2}{r^2}\right) \phi _1''(r) \notag\\
	 &&\quad +\frac{2}{r}\left(1-\frac{G M}{r}\right) \phi_1'(r)
	 -\frac{1}{a G}\phi _1(r)=0 ~,
	 \label{eq:phi1-first}
\end{eqnarray}
which quantifies the deviation $s\phi_{1}(r)$ from general relativity.
Unfortunately, even this equation is not that cooperative.
At large $r$, however, which is equivalent to neglecting $M$ and $Q$,
one gets rid of the diverging term to stay with the converging Yukawa tail
$\displaystyle{\sim \frac{1}{r}e^{-\frac{r}{\sqrt{aG}}}}$.
Consequently, once $M$ and $Q$ are re-introduced, we expect the
solution to be of the form
\begin{equation}
	\phi _1 (r) = \frac{f(r)}{r}e^{-\frac{r}{\sqrt{aG}}}~.
\end{equation}
Clearly, the equation for $f(r)$ is still quite complicated, but it becomes
manageable upon keeping only the leading terms at large $r$, that is
\begin{equation}
	\sqrt{a G} f''(r) -2f'(r) -\frac{2 G M}{\sqrt{a G}r}f (r)\simeq 0~.
\end{equation}
The solution of this equation involves the Hypergeometric and the
so-called MeijerG functions.
It so happens, however, that both these functions exhibit identical
divergent behavior. 
In turn, one can always find a converging linear combination, with the latter
being proportional to $\displaystyle{r^{-\frac{G M}{\sqrt{a G}}}}$.
To be more specific,
\begin{equation}
	\phi_1(r)\simeq
	\frac{e^{-\frac{r}{\sqrt{a G}}} }{r^{1+\frac{G M }{\sqrt{a G}}}}~.
	\label{phi1}
\end{equation}

With this in hands, we proceed to the linear differential equation for
$L_1(r)$, namely
\begin{eqnarray}
	&&L_1(r)+r L_1'(r) \notag\\
	&&=-\frac{G}{r} \left(Q^2-Mr\right)\phi_1'(r)
	+\left( \frac{G Q^2}{r^2}-\frac{r^2}{a G}\right)\phi _1(r)~.
	\label{eq:L}
\end{eqnarray}
The solution of this equation is a sum of several Gamma functions which,
at the large $r$ limit, is well approximated by the expression
\begin{equation}
	L_1(r)\simeq \frac{1}{\sqrt{a G}}
	\frac{e^{-\frac{r}{\sqrt{a G}}}}{r^{\frac{ G M}{\sqrt{a G}}}}~.
	\label{L1}
\end{equation}
By the same token, one can also calculate
\begin{equation}
	N_1(r)\simeq 
	-\frac{e^{-\frac{r}{\sqrt{a G}}}}{r^{1+\frac{ G M}{\sqrt{a G}}}}~.
	\label{N1}
\end{equation}

\section{Preliminary numerical insight}
\begin{figure}[h]
	\includegraphics[scale=0.6]{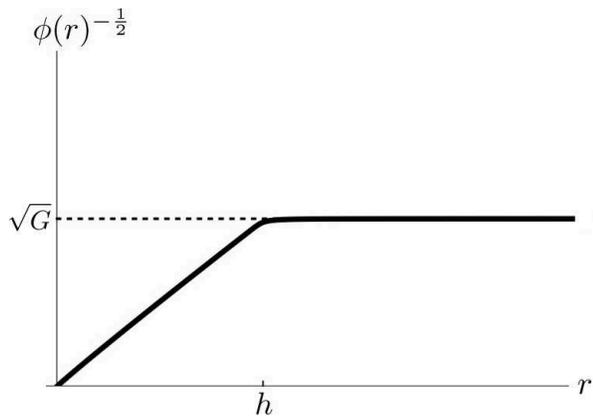}
	\caption{A generic scalar field configuration.
	As $s\rightarrow +0$, general relativity is recovered at the
	exterior region, but is spontaneously violated in the inner core.}
	\label{Fig1}
\end{figure}

\begin{figure}[h]
	\includegraphics[scale=0.6]{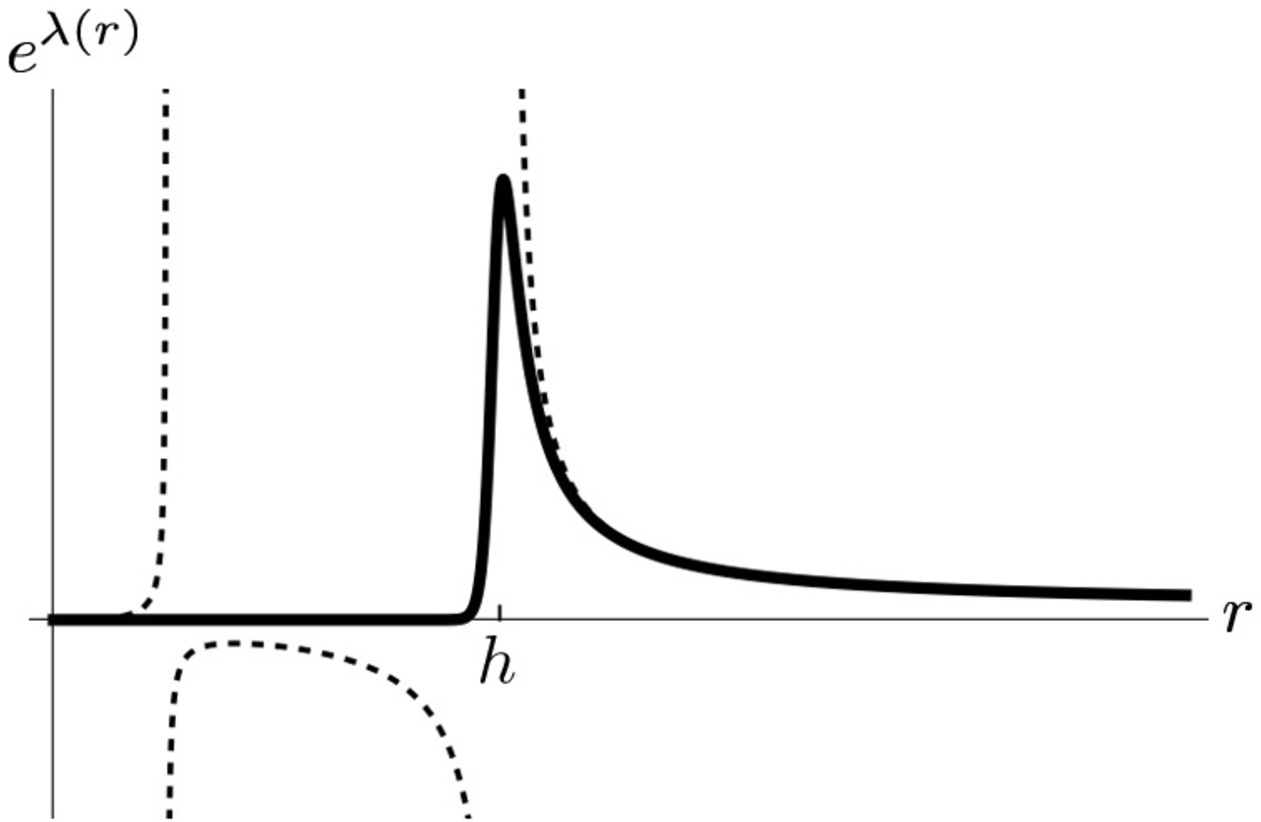}
	\caption{A generic $e^{\lambda(r)}$ plot.
	Whereas the exterior RN is recovered at the $s\rightarrow +0$ limit,
	the overall configuration conceptually differs from the full $s=0$
	RN solution (dashed line).}
	\label{Fig2}
\end{figure}

\begin{figure}[h]
	\includegraphics[scale=0.6]{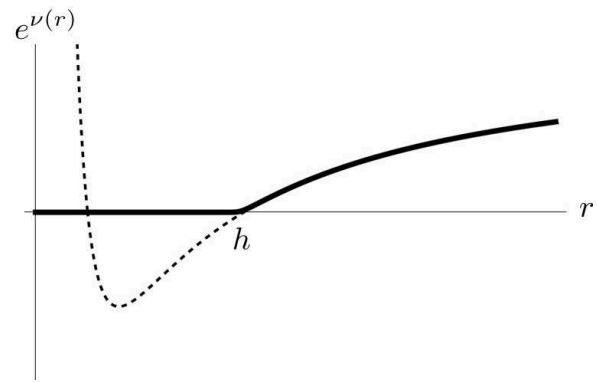}
	\caption{A generic $e^{\nu(r)}$ plot.
	Whereas the exterior RN is recovered at the $s\rightarrow +0$ limit,
	the overall configuration conceptually differs from the full $s=0$
	RN solution (dashed line).}
	\label{Fig3}
\end{figure}

On pedagogical grounds, to have a glimpse at the new physics offered by
spontaneously induced general relativity, we first plot numerical graphs of
the various functions involved.
Starting at some large enough distance
\begin{equation}
	r_{max}>>GM, \sqrt{G}Q, \sqrt{aG}~,
\end{equation}
where the Reissner Nordstrom metric components and the inverse Newton
constant are supplemented by the perturbations
eqs.(\ref{phi1},\ref{L1},\ref{N1}), we run a full numerical calculation which
produces Figs.(\ref{Fig1},\ref{Fig2},\ref{Fig3}), respectively.
We do it for a positive scalar charge $s>0$, and focus our attention on
the limit $s\rightarrow +0$.
The dashed lines in these graphs depict the underlying Reissner-Nordstrom
solution.

Naively, one would expect perhaps a full recovery of general relativity
at the $s\rightarrow 0$ limit, but can already suspect the appearance
of a phase transition near the would have been outer horizon, at
\begin{equation}
	h=GM+\sqrt{G^2 M^2-G Q^2}~.
	\label{h}
\end{equation}
Serendipitously, representing a 'level crossing' effect (soon to be clarified),
\emph{the limit $s\rightarrow +0$ does not reproduce the $s=0$ solution}.
While the exterior Reissner Nordstrom solution is recovered, which
is indeed an important feature by itself, it now connects with a novel
interior core.
This new interior solution differs conceptually from the Reissner Nordstrom
interior by three characteristic features, namely
\begin{enumerate}
	\item No $t \longleftrightarrow r$ signature flip,
	\item Drastically suppressed $e^{\nu(r),\lambda(r)}$, and
	\item Locally varying effective Newton constant.
\end{enumerate}
Following the preliminary numerical insight, we now proceed to uncover
the geometry/physics of the inner core, and reveal the analytic structure
of the phase transition profile.

\section{A novel core}

\begin{figure}[h]
	\includegraphics[scale=0.6]{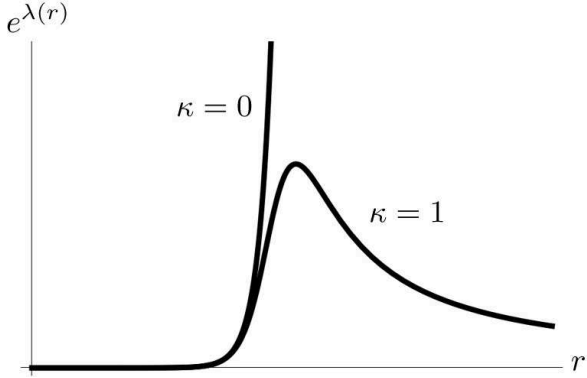}
	\caption{The suppression of $e^{\lambda(r)}$ in the inner core
	is fully captured by the $\kappa=0$ approximation.
	$\phi R$ gravity, switched on by $\kappa=1$, already exhibits
	the transition profile.}
	\label{Fig4}
\end{figure}

The key feature now is the fact that $0<e^{\lambda(r)}\ll1$ in the entire
inner core.
A closer numerical inspection reveals that all terms in the field equations
eqs.(\ref{eq:FieldEqI},\ref{eq:FieldEqIII},\ref{eq:FieldEqII}) which are proportional
to $e^{\lambda (r)}$ are practically negligible relative to the other terms.
In particular, the negligible pieces include the scalar potential terms and
the electromagnetic energy momentum contributions, thereby indicating
a universal inner structure.
The field equations take then the slimmer form
\begin{eqnarray}
	&& \phi ^{\prime\prime }-\frac{\nu'+\lambda'}{2}
	\left(\phi '+\frac{2}{r}\phi \right)=0~, \\
	&& \phi''+\frac{\nu'-\lambda'}{2}
	 \left(\phi'-\frac{2}{r}\phi\right)
	 -\frac{2 }{r^2} \left(1-\kappa e^{\lambda}\right)\phi =0~,\\
	&& \phi''  +\left(\frac{2 }{r}+\frac{\nu'
	-\lambda'}{2}\right) \phi '=0~. 
	 \label{phiApproxEq}
\end{eqnarray}
Strictly for the current approximation $\kappa=0$, but by introducing
$\kappa$ we reserve the option of searching for the roots of the transition
profile already within the $\kappa=1$ framework which corresponds to
plain (insensitive to the scalar potential) $\phi R$ gravity.
Switching $\kappa$ off and on is demonstrated in Fig.\ref{Fig4}.

The above set of scale invariant equations admit an exact analytic solution
given by
\begin{eqnarray}
	&& e^{\nu (r)}=\alpha
	\left(\frac{r}{h}\right)^{\frac{6}{\epsilon }-4}~, \notag \\
	&& e^{\lambda (r)}=\beta 
	\left(\frac{r}{h}\right)^{\frac{6}{\epsilon}-6+2\epsilon}~,
	\label{InEqs}\\
	&& \phi (r)=\gamma 
	\left(\frac{r}{h}\right)^{-2+\epsilon}~.  \notag 
\end{eqnarray}
This general solution is governed by a constant of integration
$\epsilon \neq 0$.
The self consistency of the approximation for all $r<h$ further requires 
$0<\epsilon \ll 1$.
The scale $h$, however, marking the radius of the core (where $r/h$
ceases to be a fraction), is fictitious at this stage, and can be absorbed
by re-defining the coefficients $\alpha,\beta,\gamma$.
This is a consequence of the fact that eqs.(\ref{InEqs}) are scale invariant.
It remains to be seen what actually fixes the small value of $\epsilon$,
removes the arbitrariness of $\alpha,\beta,\gamma$ (in particular
$\displaystyle{\gamma\rightarrow \frac{1}{G}}$,
as suggested by Fig.\ref{Fig1} on matching grounds), and furthermore
turns the scale $h$ into the physical quantity defined by eq.(\ref{h}).
At any rate, following our analysis so far, one may rightly suspect an
intriguing correlation between the $\epsilon\rightarrow +0$ limit
which characterizes the short distance physics, and the $s\rightarrow +0$
limit relevant for the large distance physics. 

\section{Phase transition}
Focus attention now in the neighborhood of $r=h$, where $e^{\lambda(r)}$
acquires its maximum value.
We already know, based on numerical evidence (running Fig.\ref{Fig4} for
a variety of small $\epsilon$ values, that
$e^{\lambda(h)}\sim \epsilon^{-1}$, and estimate the width of the transition
area to be $\Gamma \sim \epsilon$.
One may even suspects a universal self similar transition profile.

We probe the transition at the approximation where
\begin{equation}
	e^{\lambda (r)}\left(\phi (r)-\frac{1}{G}\right)
	\simeq 0 ~.
\end{equation}
Just inside, it is the $e^{\lambda(r)}$ factor which is highly suppressed,
whereas for $r\geq h$ a similar suppression role is played by 
$\left(\phi (r)-\frac{1}{G}\right)$.
The approximate eq.(\ref{phiApproxEq}) stays then valid at the transition 
region as well, and upon a first integration, gives rise to the (negative)
conserved quantity
\begin{equation}
	 \phi '(r)r^2e^{\frac{\lambda (r)-\nu (r)}{2}}\simeq C ~.
	 \label{conserved}    
\end{equation}
With eq.(\ref{conserved}) incorporated, setting $r\simeq h$ and
$\displaystyle{\phi(r)\simeq \frac{1}{G}}$ when appropriate, and neglecting
relatively small terms, the two remaining recast field equations read
\begin{eqnarray}
	&& \nu'-\lambda'\simeq
	\frac{2}{h}\left(1-\frac{G Q^2}{h^2}\right)e^{\lambda} ~,
	\label{pair1}\\
	&& \nu'+\lambda'\simeq
	-\frac{2}{h}\left(1-\frac{G Q^2}{h^2}\right)
	\frac{e^{\lambda}}{1
	+\frac{2h}{G C}e^{\frac{\nu-\lambda}{2}}} ~.
	\label{pair2}
\end{eqnarray}
We find it useful to introduce
\begin{equation}
	\sigma (r)=\frac{\lambda (r)-\nu (r)}{2}~,
\end{equation}
transforming the above pair of equations into
\begin{align}
	& \sigma '(r)=-\frac{1}{h}e^{\lambda (r)}
	\left(1-\frac{G  Q^2}{h^2 }\right)
	\label{eq:basic-sigma-equation1}\\
	& \frac{\sigma ''(r)}{\sigma '(r)}-\sigma '(r)
	=\frac{\sigma '(r)e^{\sigma (r)}}{\frac{2h}{C G}+e^{\sigma (r)}} 
	\label{sigma}
\end{align}
Two successive integrations bring us then to the inverse solution $r(\sigma)$
\begin{equation}
	r-\bar{r}=-\frac{C G}{2p}\left(e^{-\sigma (r)}-\frac{C G}{2h}
	\log\left(-\frac{2h}{C G} e^{-\sigma (r)}-1\right)\right)
	\label{sigma_sol} ~,
\end{equation}
with two constants of integration $p$ and ${\bar r}$ floating around.
Substituting back into eqs.(\ref{pair1},\ref{pair2}), we finally arrive at the
parametric solution
\begin{align}
	& e^{\nu (r)}=-\frac{p e^{-\sigma (r)}}{1-\frac{G  Q^2}{h^2 }}
	\left(\frac{2h}{C G}+e^{\sigma (r)}\right)~,
	\label{eq:basic-horizon-equation3} \\
	& e^{\lambda (r)}=-\frac{p e^{\sigma (r)}}{1-\frac{G  Q^2}{h^2 }}
	\left(\frac{2h}{C G}+e^{\sigma (r)}\right) ~,
	\label{eq:basic-horizon-equation2}
\end{align}
which we now attempt to connect with the approximate solutions for the
exterior and the interior regimes.
To do so, it is convenient to introduce a dimensionless variable
\begin{equation}
 	x=-\frac{2h}{C G} e^{-\sigma (r)} ~,
\end{equation}
so that
\begin{eqnarray}
	&& r-\bar{r}=\frac{h}{ p }
	\left(\frac{C G}{2 h}\right)^2     \left(x+\log (x-1)\right)~,
	\label{eq:basic-match-equation1} \notag \\
	&& e^{\nu (r)}= -\frac{p}{1-\frac{G  Q^2}{h^2 }}(1-x)~.
	\label{eq:basic-match-equation3} \\
	&& e^{\lambda (r)}=
	-\frac{p\left(\frac{2h}{C G}\right)^2}{1-\frac{G Q^2}{h^2 }}
	\left(\frac{1-x}{x^2}\right)~,
	\label{eq:basic-match-equation2}  \notag
\end{eqnarray}

\medskip
\noindent {\bf{1. Matching with the Exterior}}
\medskip

In the exterior, let $\displaystyle{x=\frac{1}{\delta}}$ for $0<\delta \ll 1$, hence
\begin{eqnarray}
	&& r-\bar{r}\simeq
	\frac{C^2 G^2}{4p h }\frac{1}{\delta} ~, \notag \\
	&& e^{\nu (r)}
	\simeq \frac{p}{1-\frac{G Q^2}{h^2}}\frac{1}{\delta }
	\simeq \frac{r-\bar{r}}{\frac{C^2 G^2}{4p^2 h }
	\left(1-\frac{G Q^2}{h^2}\right)} ~, \label{matchEx}\\
	&& e^{\lambda (r)}\simeq
	\frac{p\left(\frac{2h}{C G}\right)^2}{1-\frac{G Q^2}{h^2}}\delta
	\simeq\frac{ h}{\left(1-\frac{G Q^2}{h^2}\right)
	\left(r-\bar{r}\right)}~. \notag
\end{eqnarray}
This  can be immediately recognized as the leading expansion terms
just outside the  Reissner Nordstrom outer horizon.
In turn, up to first order corrections, we identify
\begin{eqnarray}
	& {\bar r}=h ~,& \\
	& \displaystyle{\frac{p}{C^2}=
	\frac{G^2}{4h^2p} \left(1-\frac{G Q^2}{h^2}\right)^2~.}&
\end{eqnarray}

\medskip
\noindent {\bf{2. Matching with the Interior}}
\medskip

In the interior, let $x=1+\delta$ for $0<\delta \ll 1$, so that
$\displaystyle{r-h\simeq \frac{C^2 G^2}{4p h } \log \delta}$.
Although $\log\delta$ is large, 
$\displaystyle{\frac{C^2 G^2}{4p h^2} \log \delta}$ may still be
small.
In which case,
\begin{eqnarray}
	&& \frac{r}{h}=\delta ^{\frac{C^2 G^2}{4ph^2}} ~, \notag \\
	&& e^{\nu (r)}\simeq \frac{p}{1-\frac{G Q^2}{h^2}}\delta
	\simeq \frac{p}{1-\frac{G Q^2}{h^2}}
	\left(\frac{r}{h}\right)^{\frac{4p h^2}{C^2 G^2}}~,\\
	 && e^{\lambda (r)}
	 \simeq \frac{p \frac{4h^2}{C^2 G^2}}{1-\frac{G Q^2}{h^2}}\delta
	 \simeq \frac{p \frac{4h^2}{C^2 G^2}}{1-\frac{G Q^2}{h^2}}
	 \left(\frac{r}{h}\right)^{\frac{4p h^2}{C^2 G^2}} ~. \notag
\end{eqnarray}
This set is nothing but the previously derived eqs.(\ref{InEqs}) provided
we make the identification
\begin{equation}
	\frac{\epsilon}{6} = \frac{p}{\left(1 - \frac{G Q^2}{h^2}\right)^2} ~.
\end{equation}
We can now furthermore fix the otherwise arbitrary coefficients
$\alpha,\beta,\gamma$ which enter the core metric
\begin{equation}
	\alpha (\epsilon)=\frac{1}{\beta (\epsilon)}=
	\left(1-\frac{G Q^2}{h^2}\right)\frac{\epsilon}{6}~,
	\quad \gamma (\epsilon)=\frac{1}{G} ~. 
	\label{outEqs}
\end{equation}

\medskip
\noindent {\bf{3. Self similar transition profile}}
\medskip

\begin{figure}[h]
	\includegraphics[scale=0.6]{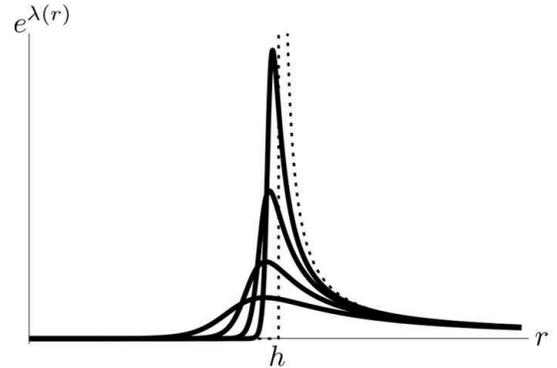}
	\caption{The in-out transition profile is plotted for a decreasing
	series of $\epsilon$ values.
	The phase transition into the exterior Reissner Nordstrom
	solution (dashed line) occurs as $\epsilon \rightarrow +0$.}
	\label{Fig5}
\end{figure}

The maximum of $e^{\lambda (r)}$, serving as the characteristic
cut-off of spontaneously induced general relativity, occurs for
$\displaystyle{e^{-\sigma}=-\frac{CG}{h}}$,
and takes the value
\begin{equation}
	e^{\lambda}_{max}
	=\frac{3}{2\epsilon \left(1-\frac{G Q^2}{h^2 }\right)}  ~.
	\label{Lmax}
\end{equation}
The typical width $\Gamma$, where $e^{\lambda (r)}$ drops to
half its maximal size, is given by
\begin{equation}
	\Gamma\simeq \frac{\epsilon}{3}h~,
\end{equation}
such that $e^{\lambda}_{max}\Gamma$ stays $\epsilon$-independent.
These features are demonstrated in Fig.\ref{Fig5} for a variety of
$\epsilon$ values (the dashed line represents the
$\epsilon\rightarrow +0$ limit).

Owing to the fact that its roots are located in the underlying $\phi R$
gravity, the transition profile turns out to be insensitive to the terms
involving the scalar potential, and exhibits a remarkable self similarity
feature.
This is expressed by the fact that $\epsilon\rightarrow
k\epsilon$ only causes scale changes
\begin{align}
	& e^{\lambda}\rightarrow k^{-1}e^{\lambda} ~,\\
	& r-h\rightarrow k(r-h)~.
\end{align}

\medskip
\noindent {\bf{4. The $\epsilon \longleftrightarrow s$ interplay}}
\medskip

$\epsilon$ parametrizes the short distance geometry, whereas the
scalar charge $s$ parametrizes the long distance perturbation around
the Reissner Nordstrom background.
The remarkable correlation between the small-$\epsilon$ and the
small-$s$ limits has already been qualitatively established, but it
seems impossible to derive the exact $\epsilon (s)$ relation, and and
at this stage, has only been obtained numerically.
This can be done e.g. by plotting
$\frac{12}{r}\left (\nu'(r)+\lambda'(r)\right)\simeq \epsilon$ at short
distances, or alternatively by extracting
$(e^{\lambda}_{max})^{-1}\sim \epsilon$ at the transition region.

\begin{figure}[h]
	\includegraphics[scale=0.6]{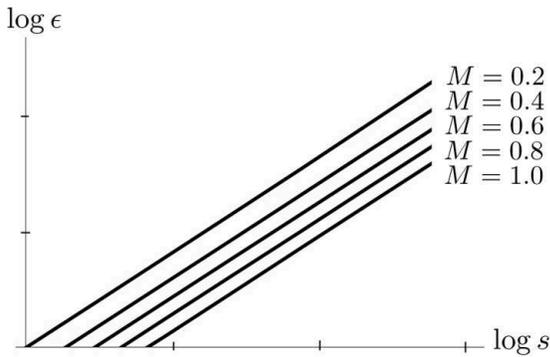}
	\caption{The $\epsilon \longleftrightarrow s$ interplay:
	The linearity of $\epsilon(s)$ is demonstrated for
	various values of $M$ (in units of $\sqrt{G/a}$).}
	\label{Fig6}
\end{figure}

First, one can numerically verify that $\epsilon (s)$ is practically
$a$-independent, at least for large enough $a$'s.
Thus, holding $a$ fixed, we then plot $\epsilon (s)$ for various values
of $M$ (and momentarily keep $Q=0$). 
The results are summarized in Fig.\ref{Fig6}.

From the slope we deduce that, for large $a$, $\epsilon$ is
proportional to $s$, and that the proportionality factor is
$M$-dependent.
Taking into account the structure of the $s$ term which enters
the approximation eq.(\ref{phi1}), and invoking continuity
arguments (next order necessarily included) at $r\simeq h$,
we elegantly fit the numerical graphs by the empirical formula
\begin{equation}
	\epsilon\simeq \frac{4 e^{-\frac{h}{\sqrt{aG}}}s}
	{\left( 1-\frac{GQ^2}{h^2}\right)
	h^{1+\frac{GM}{\sqrt{aG}}}}~.
\end{equation}

\section{Core geometry}

Altogether, the core metric is well approximated by
\begin{equation}
	ds^2_{in}\simeq-\alpha(\epsilon)
	\left(\frac{r}{h}\right)^{\frac{6}{\epsilon }-4}dt^2+
	\frac{1}{\alpha(\epsilon)}
	\left(\frac{r}{h}\right)^{\frac{6}{\epsilon}-6+2\epsilon}dr^2+
	r^2 d\Omega^2~,
	\label{InMetric}
\end{equation}
where $\alpha(\epsilon)$ is given explicitly by eq.(\ref{outEqs}). 
This metric is accompanied by the associated scalar field
\begin{equation}
	\phi(r)\simeq
	\frac{1}{G} \left(\frac{r}{h}\right)^{-2+\epsilon} ~.
\end{equation}
As $\epsilon \rightarrow +0$, the metric connects with the
perturbed exterior Reissner Nordstrom spacetime, and the scalar
field approaches its general relativistic VEV.

\medskip
\noindent {\bf{1. Vanishing volume}}
\medskip

\begin{figure}[h]
	\includegraphics[scale=0.6]{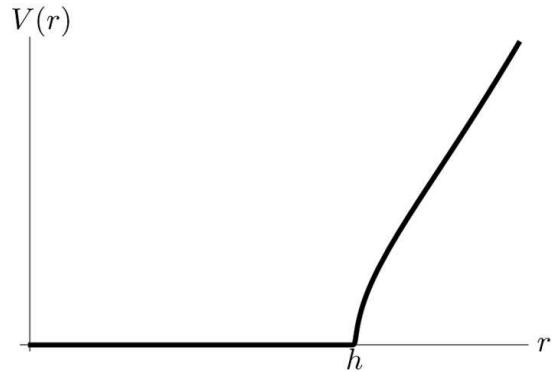}
	\caption{The invariant volume $V(r)$:
	Every inner concentric sphere of finite surface area $4\pi r^2$
	exhibits a vanishingly small volume.}
	\label{Fig7}
\end{figure}

The first geometrical quantity to calculate is the invariant spatial
volume $V(r)$ associated with a sphere of circumferential radius $r$.
Doing it in the interior core, one approaches a vanishingly
small volume at the $\epsilon \rightarrow +0$ limit, namely
\begin{align}
	& V(r)=4\pi\int_0^r e^{\frac{1}{2}\lambda(r)}r^2~dr \notag \\
	& \quad \simeq \sqrt{\frac{6\epsilon}
	{1-\frac{G Q^2}{h^2}}}
	\left(\frac{r}{h}\right)^{\frac{3}{\epsilon }}
	\frac{4\pi}{3} h^3 ~.
\end{align}
This is to be fully contrasted with the corresponding finite surface area
\begin{equation}
	A(r)=4\pi r^2~.
\end{equation}
The $V(r)$ plot, depicted in Fig.\ref{Fig7}, gives us a simple answer
why is the black hole volume physically irrelevant (unlike in any other
system, the black hole entropy is proportional to the horizon surface
area).

\medskip
\noindent {\bf{2. Light cone structure}}
\medskip

Radial null geodesics at $r=h+\delta r$, just outside the would have
been horizon, obey
\begin{equation}
	\frac{dr}{dt}=\pm
	\left(1-\frac{G Q^2}{h^2}\right)\frac{\delta r}{h} ~.
	\label{GeoOut}
\end{equation}
Inside the core, the formula transforms into
\begin{equation}
	\frac{dr}{dt}=\pm
	\left(1-\frac{G Q^2}{h^2}\right)\frac{\epsilon r}{6h}~,
	\label{GeoIn}
\end{equation}
where the exact role played by $\delta r$ in eq.(\ref{GeoOut}) is being
taken by $\frac{1}{6}\epsilon r$ in eq.(\ref{GeoIn}).
In particular, as far as an observer at asymptotic distances is concerned,
a light ray sent from some $r_0<h$ inwards will never reach the origin
even for a finite $\epsilon$, as can be seen from
\begin{equation}
	r(t)=r_0 e^{-t/\tilde{t}},
	\quad \tilde{t}=\frac{6h}{\epsilon
	\left(1-\frac{GQ^2}{h^2}\right)}~.
\end{equation}
In other words, \emph{the entire core resembles a 'near horizon' territory},
and as $\epsilon \rightarrow +0$, it looks from the outside as an apparently
'frozen world' \cite{Paddy}.
In many respects, the role played by the outer Reissner Nordstrom event
horizon gets now shifted to the origin.

\medskip
\noindent {\bf{3. Constant surface gravity}}
\medskip

To get a deeper clue about what is going on, we calculate the surface gravity
function
\begin{equation}
	\kappa (r)=\frac{\nu'(r)}{2}e^{\frac{\nu(r)-\lambda(r)}{2}}
\end{equation}
inside the core, and find
\begin{equation}
	\kappa (r)\simeq \left(1-\frac{2}{3} \epsilon \right)
	\frac{1-\frac{G Q^2}{h^2} }{2h }\left(\frac{h}{r}\right)^{\epsilon }
	\rightarrow \kappa ~.
	\label{kappa}
\end{equation}
Not only do we face a constant surface gravity core, all the way from
$r\leq h$ to the origin at $r=0$, but its value is immediately identified
as $2\pi$ the Hawking temperature
\begin{equation}
	\boxed{\kappa = \frac{1}{2h}\left(1-\frac{G Q^2}{h^2}\right)}~.
	\label{frequency}
\end{equation}
This will certainly have far reaching consequences (soon to be revealed)
on the Komar mass and the associated thermodynamics.
 
\medskip
\noindent {\bf{4. Singularity issues?}}
\medskip

When approaching the origin, it is convenient to invoke the proper length
coordinate
\begin{equation}
	\eta(r) = \frac{h\left(\frac{r}{h}\right)^{-2+\frac{3}
	{\epsilon }+\epsilon }}{\left(\frac{3}
	{\epsilon }-2+\epsilon \right)
	\sqrt{\left(1-\frac{G Q^2}{h^2}\right)\frac{\epsilon }{6}}}~,
\end{equation}
and expand the inner metric, up to ${\cal O}(\epsilon)$ pieces,
to expose the Rindler structure of the $R_{2}$ sub-metric
\begin{equation}
	ds_{in}^2\simeq
	- \kappa^2 \eta^2 dt^2+d\eta^2
	+\left(\frac{3\kappa \eta^2}{\epsilon h}\right)^{\frac{3}{\epsilon}}
	h^2 d\Omega^2~.
	\label{Rindler}
\end{equation}
The recovery the exact Hawking's imaginary time periodicity, which first of all
reassures the accuracy of the transition profile eq.(\ref{eq:basic-match-equation2}),
is regarded as the anchor connecting us to black hole thermodynamics.
Notice, however, that \emph{unlike in the original Reissner Nordstrom
case, the Euclidean origin corresponds now to the center
of spherical symmetry $r=0$ rather than to $r=h$}.

A word of caution is in order.
Inside the core, the the various scalars are well approximated by
\begin{eqnarray}
	&& R= -\frac{2}{r(\eta)^2} ~,  \\
	&& R^{\mu\nu}R_{\mu\nu}
	\simeq\frac{8\epsilon^2}{9 \eta^4}
	\left(1-\frac{G Q^2}{h^2}\right)^2
	+\left[ \frac{2}{r^4 (\eta)} \right]  ~,  \\
	&& R^{\mu\nu\lambda\sigma} R_{\mu\nu\lambda\sigma}
	\simeq\frac{16\epsilon^2}{9 \eta^4}
	\left(1-\frac{G Q^2}{h^2}\right)^2
	+\left[ \frac{4}{r^4 (\eta)} \right] ~,
\end{eqnarray}
where the square parentheses $\left[...\right]$ denote relatively small
terms which survive the $\epsilon \rightarrow 0$ limit.
Reflecting the $\displaystyle{\frac{\epsilon}{\eta^2}}$ ratio, the singularity 
analysis bifurcates:

\medskip
\noindent (i) Clearly, for any finite $\eta$, as small as desired,
the limit $\epsilon\rightarrow 0$ is \emph{regular}.
The Rindler sub-metric gets then multiplied by a 2-sphere of constant
radius $h$, and consistently, the Kretschmann curvature approaches
the Reissner Nordstrom horizon value of $\displaystyle{\frac{4}{h^{4}}}$.

\noindent (ii) However, for any finite $\epsilon$, as small as desired,
the limit $\eta\rightarrow 0$ is \emph{singular}.
Whereas the pseudo-horizon does provide some protection
from the singularity (e.g. it takes an infinite amount of time
for light from the singularity to reach any external observer),
an observer willing to wait long enough will see unbounded
high curvature.
Such a behavior is far worse than that of the Reissner Nordstrom
solution, and constitutes a severe problem.
It may be that a more complicated Lagrangian could alleviate this
behavior, or else that quantum effects could eventually cure it.

The point is, however, that the parameter $\epsilon$ and the
coordinate $\eta$ cannot really be treated on the same footing.
Any given metric must first of all be specified by its parameters,
and only then can it serve the whole range of coordinates.
In turn, the right order is to first let $\epsilon\rightarrow 0$,
and only then approach the $\eta\rightarrow 0$ origin. 
In other words, using momentarily a two variable language,
$\epsilon$ must tend to zero faster than $\eta^2$ (we return
to this point in the Kruskal analysis).
This argument is supposed to solve the above dilemma by
choosing the first option.

\medskip
\noindent {\bf{5. Kruskal structure at the origin}}
\medskip

Another view on the geometry surrounding the origin is provided
by means of the Kruskal-Szekeres coordinate transformation
\begin{equation}
	u=f(r)\cosh \omega t~,~v=f(r)\sinh \omega t~,
\end{equation}
which by choosing
$\displaystyle{\frac{f'(r)}{\omega f(r)}= e^{\frac{\lambda (r)-\nu (r)}{2}}}$ 
gives rise to a metric of the form
\begin{equation}
	ds_{in}^2=
	K^2 (r)\left(-dv^2+du^2\right)+r^2 d\Omega^2~.
\end{equation}
In our case, we find
\begin{equation}
	\log f(r)\simeq \frac{6\omega h}{ \left(1-\frac{G Q^2}{h^2}\right)}
	\frac{\left(\frac{r}{h}\right)^{\epsilon }}{\epsilon ^2}~,
\end{equation}
and the crucial point has to do with the $\epsilon$-expansion
\begin{equation}
	\left(\frac{r}{h}\right)^{\epsilon }
	=1 + \epsilon \log \frac{r}{h}
	+\frac{1}{2}\left(\epsilon \log \frac{r}{h}\right)^2+...~.
\end{equation}
As was emphasized earlier, a small parameter and a small coordinate
cannot be treated on equal footing.
The parameter $\epsilon$ which specifies the metric must tend to zero
faster than any function of the coordinates,
say $\left(\log \frac{r}{h}\right)^{-1}$ in this case.
The corresponding Kruskal scale function $K(r)$ is then given by
\begin{equation}
	K(r)\simeq\left(\frac{r}{h}\right)^{\displaystyle{\frac{3}{\epsilon}
	\left(1-\frac{2 \omega  h}{1-\frac{G Q^2}{h^2}}\right)}}~,
	\label{Kfunction}
\end{equation}
which consistently singles out the Hawking imaginary time periodicity
eq.(\ref{frequency}) on the grounds of defusing the conic singularity
at the origin.

\medskip
\noindent {\bf{5. Frame independence}}
\medskip

Although the Jordan frame is the physical one in the hereby discussed theory,
the Einstein frame is still of interest.
In 4-dimensions, the transition is established by substituting
\begin{equation}
	g_{\mu\nu}=\phi^{-1}\bar{g}_{\mu\nu}~.
\end{equation}
By an accompanying change of variables, namely by
\begin{equation}
	\rho=r \left(\frac{r}{h}\right)^{1-\frac{\epsilon}{2}},
\end{equation}
the resulting metric $d\bar{s}^2_{in}$ takes the form
\begin{equation}
	d\bar{s}^2_{in}=
	-\alpha (\epsilon )\left(\frac{\rho }{h}\right)^{\frac{3}{\epsilon }
	-\frac{1}{4}}dt^2+\frac{1}{4\alpha (\epsilon )}
	\left(\frac{\rho }{h}\right)^{3\left(\frac{1}{\epsilon }
	-\frac{3}{4}\right)}d\rho^2 +\rho ^2 d\Omega ^2 ~,
\end{equation}
to be compared with eq.(\ref{InMetric}).
A closer inspection reveals that all physical conclusions remain intact,
in particular the forthcoming formula of the Komar mass.

\section{Non-singular Komar Mass}

\begin{figure}[h]
	\includegraphics[scale=0.6]{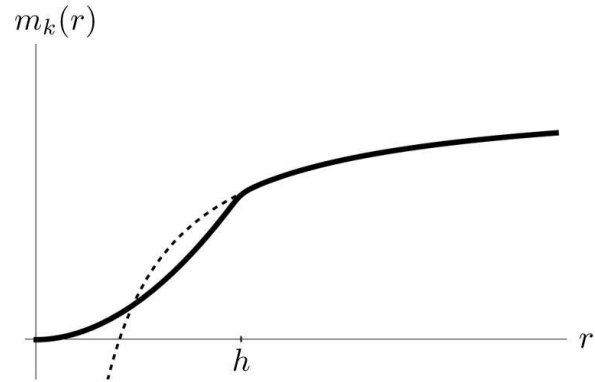}
	\caption{Unlike in the Reissnner Nordstrom case (dashed
	line), the Komar mass $m_{K}(r)$ is non-singular at the origin,
	and furthermore obeys the positive energy condition.}
	\label{Fig8}
\end{figure}

General relativity does not offer a unique definition for the
term mass.
The ADM mass, for example, only makes sense globally, at
asymptotically flat spatial infinity.
But in the presence of a timelike Killing vector, like in the present
case, it is the Komar mass which becomes a tenable choice.
Invoking Stoke's theorem and performing the angular integration
for a static spherically symmetric metric, the Komar mass \cite{Komar}
becomes proportional to the surface gravity, and is simply given by
\begin{equation}
	m_K (r)=\kappa(r) \frac{r^2}{G}~.
\end{equation}
In the exterior region, it exhibits the familiar classical formula
\begin{equation}
	m_K^{out} (r)\simeq M-\frac{Q^2}{r}~.
\end{equation}
The constant mass term (the mass sources are solely in the interior) is
accompanied by the familiar decreasing electromagnetic contribution.
For the Reissner Nordstrom geometry, this results hold everywhere,
including in the interior, but this is definitely not the case here.
To be specific, in our interior region, we derive
\begin{equation}
	m_K^{in}\simeq
	\left(1-\frac{2}{3} \epsilon \right)\left(\frac{h}{r}\right)^{\epsilon}
	\frac{\kappa r^2}{G}~,
\end{equation}
and immediately appreciate its $\epsilon \rightarrow 0$ limit
\begin{equation}
	\boxed{m_K^{in}(r)\rightarrow
	\left( M-\frac{Q^2}{h} \right)\frac{r^2}{h^2}}~.
	\label{Komarcore}
\end{equation}
Every concentric inner sphere of invariant surface area $A(r)=4\pi r^2$
carries a geometric fraction $\displaystyle{\frac{A(r)}{A(h)}}$ of the total
Komar mass enclosed by the would have been Reissner Nordstrom outer
horizon.
The Komar mass function is plotted in Fig.\ref{Fig8}.
It exhibits two truly exceptional features which the general relativistic
Reissner Nordstrom metric simply falls short to provide. 
To be specific, $m_K (r)$ is now
\begin{itemize}
	\item Non-singular at the origin.
	\item Non-negative definite.
\end{itemize}
The Komar mass is \emph{universally distributed} all over the core,
and the positive energy condition is automatically respected.

A note is now in order.
The so-called material energy is a tenable alternative to the Komar mass.
Following Weinberg \cite{Weinberg}, it is the integration of the
energy as measured in a locally inertial frame.
Technically speaking, one supplements the naive (non-covariant)
mass formula by the missing $\sqrt{-g_{tt}g_{rr}}$ factor, to give
\begin{equation}
	M_{W}(r)=\frac{1}{G} 
	\int_0^{r} e^{\frac{1}{2}(\nu+\lambda)}
	\left(  1+e^{-\lambda}(\lambda^{\prime}r-1)
	\right)dr ~,
	\label{Weinberg}
\end{equation}
for a spherically symmetric metric.
With regard to the present case, it seems that  the two mass definitions,
Komar mass and Weinberg's material energy, are distinguished from
each other only by their different ${\cal O}(\epsilon)$ corrections.

\section{Holographic entropy packing}

The geometric anchor connecting us to black hole thermodynamics
is the imaginary time periodicity of the Euclidean manifold (or alternatively,
the Kruskal $\omega$ parameter which characterizes the Lorentzian
manifold), which underlies the Hawking temperature 
\begin{equation}
	T=\frac{\kappa}{2\pi}=
	\frac{1}{4\pi h}\left(1-\frac{GQ^2}{h^2}\right)~.
\end{equation}
$T=T_{\infty}$  is the temperature \emph{at infinity} associated with the thermal
state of the field theory which lives on the black hole background
The striking feature is that, unlike in conventional black hole physics,
the exact Hawking periodicity has been recovered in the present theory
by defusing the conic defect at the origin, rather that at the event horizon.
A variety of related features which include
\begin{itemize}
	\item 'Near horizon' like light cone structure eq.(\ref{GeoIn}),
	\item Equi surface gravity eq.(\ref{kappa}),
	\item Rindler structure eq.(\ref{Rindler}),
	\item Kruskal structure eq.(\ref{Kfunction}),
	\item Universal Komar mass eq.(\ref{Komarcore}),
\end{itemize}
all point out towards non-trivial physics associated with the black hole
interior.

Starting from the Smarr formula \cite{Smarr}
\begin{equation}
	 m_{K}(h)=\frac{\kappa}{4\pi} A(h)~,
	 \label{Smarr}
\end{equation}
or more precisely, from its thermodynamic oriented formulation
\begin{equation}
	  M-\frac{Q^2}{h}=2TS~,
	   \label{Smarrh}
\end{equation}
one first confronts the $\{M,Q\}$ black hole with its
$\{M+\Delta M, Q+\Delta Q\}$ extension, to find
\begin{equation}
	 \Delta M-\frac{Q}{h} \Delta Q =T \Delta S ~.
	 \label{DeltaSmarrh}
\end{equation}
The result, as is well known, is the 1st law of charged black hole
thermodynamics.
Next we multiply eq.(\ref{Smarrh}) by the $r^2/h^2$ ratio, to
obtain a meaningful formula for an inner sphere of circumferential
radius $r\leq h$
\begin{equation}
	m_{K}^{in}(r)=\frac{\kappa}{4\pi} A(r)~,
\end{equation}
or equivalently
\begin{equation}
	\left(M-\frac{Q^2}{h}\right)\frac{r^2}{h^2}
	=2Tֿ\left(S\frac{r^2}{h^2}\right)~,
\end{equation}
and attempt to understand its significance.
Note that an analogous formula simply does not exist for the interior
of ordinary black holes.
In some respects, as could already been inferred from the light
cone structure, and from the other features on the list at the
beginning of this section, the inner spherical surface resembles
in some respects a \emph{maximally stretched horizon}.
Consequently, it becomes meaningful to ask what portion $S(r)$
of the total entropy $S\equiv S(h)$ is stored within an arbitrary
inner sphere of a finite surface
area $A(r)=4\pi r^{2}$ (and most importantly, of a vanishing
invariant volume $V(r) \rightarrow 0$) which hosts a Komar mass
$M_K(r)$?
Already at this stage, one can already deduce that
\begin{equation}
	\boxed{S(r)=S\frac{r^2}{h^2}=\frac{\pi r^2}{G}}~,
	\label{universalS}
\end{equation}
and appreciate the emerging purely geometrical  't Hooft-Susskind
universal entropy bound eq.(\ref{bound}), and the fact that the bound
is locally saturated.

\begin{figure}[h]
	\includegraphics[scale=0.5]{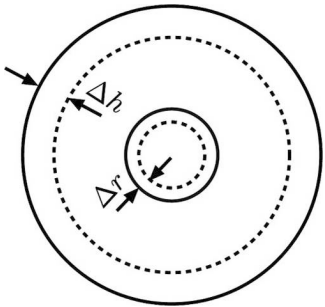}
	\caption{The $\{M+\Delta M,Q+\Delta Q\}$ configuration
	(solid circles) is a linearly stretched version of the $\{M,Q\}$
	configuration (dashed circles). As $h \rightarrow h+\Delta h$,
	each point at a circumferential radius $r$ gets radially shifted
	by an amount $\Delta r=\displaystyle{\frac{r}{h}\Delta h}$.}
	\label{Fig9}
\end{figure}

It remains, however, to figure out how exactly does the core
configuration change when supplementing the pair $\{M, Q\}$ by
tiny amounts $\{\Delta M,\Delta Q\}$, respectively?
The crucial point to notice then is that, at the
$\epsilon \rightarrow 0$ limit, \emph{the resulting configuration
appears to be nothing but a linearly stretched version of the former
configuration}.
Exactly in the same way that an ordinary black hole changes it
size once $M,Q$ get shifted, meaning $h \rightarrow h+\Delta h$
accordingly, any infinitely thin concentric layer of radius $r$ is puffed
up to a new radius $r \rightarrow r+\Delta r$ (see illustration in
Fig.\ref{Fig9}), in such a way that
\begin{equation}
	\Delta \left(\frac{r}{h}\right)=0 ~~
	\Longrightarrow ~ ~\Delta r=\frac{r}{h} \Delta h~.
	\label{Delta}
\end{equation}
This is a reflection of the fact that, as $\epsilon \rightarrow 0$,
the one and only length scale floating around is $h$.
Altogether, in analogy with eq.(\ref{DeltaSmarrh}), and subject to
the geometric rule eq.(\ref{Delta}), we have
\begin{equation}
	\Delta \left( M\frac{r^2}{h^2} \right)
	-\frac{1}{h}\left( Q\frac{r}{h} \right)\Delta \left( Q\frac{r}{h} \right)
	=T\Delta S(r)~.
\end{equation}

The emerging entropy packing profile turns out to be
(i) Locally holographic, i.e. exhibits proportionality to
$A(r)$ for every $r\leq h$, and
(ii)  $M,Q$-independent, and hence universal.
The overall picture is then of an onion-like entropy packing model \cite{DG}.
The entropy of any inner sphere is maximally packed, and
unaffected by the outer layers.
In particular, any additional piece of entropy is maximally packed
on its own external layer, with $M,Q$ as well as $M_K (r)$ being
adjusted accordingly.

An interesting point has to do with the entropy to energy ratio,
which is $r$-independent in our case, but is never smaller than
$2\pi r$
\begin{equation}
	\frac{S(r)}{M_{K}(r)}=\frac{S(h)}{M_{K}(h)}
	=\frac{\pi  h^2}{G\left(M-\frac{Q^2}{h}\right)} 
	\geq 2\pi r~.
	\label{ratio}
\end{equation}
This is in apparent violation of Bekenstein's universal entropy
bound \cite{Bbound}.
The reason seems to be the following.
Whereas the entropy $S(r)$ of some inner sphere is universal,
the associated Komar mass $M_{K}(r)\sim M^{-1}$ is affected
by the total mass $M$ of the whole system.
Admittedly, Bekenstein's universal bound is relevant
\cite{Bviolation} only for weakly self gravitating isolated physical
systems, and for these it is a much stronger bound than the
holographic one.

The various scalings involved may suggest that the holographic
entropy packing is indeed a matter of interpretation.
To be more explicit, let us examine the issue from the point of view
of a physicist who is convinced that general relativity is the
fundamental theory of gravity, and therefore is totally unaware
of its hereby advocated spontaneously induced nature.
Such a physicist (not to be confused with an Einstein frame
observer whose metric is $\phi^{-1}g_{\mu\nu}$ rather the
$g_{\mu\nu}$) would recast the underlying field  equations
into their basic Einstein form
${\cal R}_{\mu\nu}-\frac{1}{2}g_{\mu\nu}{\cal R}=
8\pi G{\cal T}_{\mu\nu}^{eff}$, moving all terms and factors, save
for the Einstein tensor itself, to the r.h.s., thereby defining
an effective energy/momentum tensor ${\cal T}_{\mu\nu}^{eff}$.
In particular, inside the core, the dynamical Newton constant is
given by
\begin{equation}
	G_{in}(r)=\frac{1}{\phi(r)}=G\frac{r^{2}}{h^{2}}~,
\end{equation}
but our 'general relativistic' observer still insists on it being $G$,
which requires from his side the effective replacement 
\begin{equation}
	G \mapsto G\frac{h^2}{r^2}~.
\end{equation}
The Hawking temperature, on the other hand, is defined at
asymptotic distances, and thus, is fully respected by our 'naive'
observer, that is
\begin{equation}
	T\mapsto T~.
\end{equation}
But this cannot be the case, unless of course
\begin{equation}
	GM\mapsto GM~,~GQ^{2}\mapsto GQ^{2}
	~~\Longrightarrow ~~ h\mapsto h~.
\end{equation}
In turn, fully consistent with our analysis, the following
counter replacements are in order
\begin{equation}
	M\mapsto M\frac{r^{2}}{h^{2}}~,~
	Q\mapsto Q\frac{r}{h}~.
\end{equation}
All the above nicely converge now back into
\begin{equation}
	S=\frac{\pi h^{2}}{G}\mapsto
	\frac{\pi r^{2}}{G}~,
\end{equation}
which completes the interpretation of an observer ignorant of the
local variations of the Newton constant inside the core.

\section{Summary and outlook}
It has come as a big surprise that spontaneously induced general
relativity does not always admit a full general relativistic limit.
Such an intriguing possibility is demonstrated in this paper at the
charged black hole level, where the exterior general relativistic
Reissner-Nordstrom solution connects with a novel holographic
interior; the phase transition takes then place precisely at the would
have been outer event horizon.
The new physics associated with the inner core has been discussed
in some details, with our main result being the local realization of the 
't Hooft-Susskind-Bousso holographic principle (the holographic
bound, as we recall, is not applicable inside ordinary black holes).
Notably, this is achieved  without invoking string theory and/or the
AdS/CFT correspondence.
Our results are not sensitive to the exact shape of the scalar
potential, thus leaving the door open a more general class of
$f(R)$ gravity models, and will only suffer minor modifications
upon introducing an optional Brans-Dicke kinetic term.

The emerging maximal entropy packing mechanism sheds new
light on how information is stored within a black hole.
The interior core, resembling now a maximally stretched horizon,
is not a 'boring' place any more (at least in the sense discussed in
the introduction), but has started functioning as Nature's ultimate
information storage.
Rather than envision bits of information evenly spread solely
on the horizon surface or in its vicinity, a bit per Planck area,
they are now universally and holographically spread in the
whole black hole interior.
Rather than tiling the horizon by Planck area patches, the traditional
way it is being done in quantum black hole models, the present work
suggests the alternative of filling up the interior with (say) light
sheet unit intervals.
The overall picture is then of an onion-like entropy packing shell
model.
Reflecting our main formula eq.(\ref{universalS}), the entropy of
any inner sphere, being geometric in nature, is maximally packed
and unaffected by the outer layers.
Any additional entropy is maximally packed on its own external
layer, with the overall mass and charge, as well as the intimately
related Komar mass distribution eq.(\ref{Komarcore}), being adjusted
accordingly.
Needless to say, exactly the same structure is expected to hold
once the cosmological constant and/or angular momentum enter
the game.

A final speculation concerning the value of $\epsilon$, the
dimensionless number which  parametrizes the deviation from
general relativity, is irresistible.
Classically, with general relativity so well established,
$\epsilon \rightarrow +0$ is indeed the limit to study.
However, having quantum mechanics in mind, and appreciating
the fact that the singularity at the origin will eventually be
disarmed quantum mechanically, it is quite appealing to imagine
a very small yet a finite $\epsilon$.
For example, the invariant width of the transition region may be
fixed by the Planck length, namely
\begin{equation}
	\sqrt{\epsilon}h\simeq \ell_{Pl}~.
\end{equation}

\acknowledgments{It is a pleasure to thank Ilya Gurwich,
the co-author of the papers preceding this work, for valuable
discussions.
Special thanks to BGU president Prof. Rivka Carmi for the
kind support.}

\end{document}